\documentclass[preprint,12pt]{elsarticle}
\usepackage{epsfig}
\usepackage{lineno}
\usepackage{rotating}
\usepackage[usenames]{color}
\usepackage{multirow}
\usepackage{longtable}
\usepackage{lscape}

\journal{Nuclear Instruments and Methods A}

\begin{document}
\begin{frontmatter}

\title{Growth and characterization of a Li$_2$Mg$_2$(MoO$_4$)$_3$ scintillating bolometer}

\author[KINR,CSNSM]{F.A.~Danevich\corref{cor1}}
\cortext[cor1]{Corresponding author. E-mail:
fedor.danevich@csnsm.in2p3.fr}
\author[KNU]{V.Ya.~Degoda}
\author[KNU]{L.L.~Dulger}
\author[CSNSM]{L.~Dumoulin}
\author[CSNSM,DISAT]{A.~Giuliani}
\author[CSNSM]{P.~de~Marcillac}
\author[CSNSM]{S.~Marnieros}
\author[CEA]{C.~Nones}
\author[CSNSM]{V.~Novati}
\author[CSNSM]{E.~Olivieri}
\author[NIIC]{A.A.~Pavlyuk}
\author[KINR,CSNSM]{D.V.~Poda}
\author[NIIC]{V.A.~Trifonov}
\author[NIIC]{I.V.~Yushina}
\author[CEA]{A.S.~Zolotarova}

 \address[KINR]{Institute for Nuclear Research, 03028 Kyiv, Ukraine}
 \address[CSNSM]{CSNSM, Univ. Paris-Sud, CNRS/IN2P3, Universit\'{e} Paris-Saclay, 91405 Orsay, France}
 \address[KNU]{Kyiv National Taras Shevchenko University, 03127 Kyiv, Ukraine}
 \address[DISAT]{DISAT, Universit\`a dell'Insubria, 22100 Como, Italy}
 \address[CEA]{IRFU, CEA, Universit\'{e} Paris-Saclay, F-91191 Gif-sur-Yvette, France}
 \address[NIIC]{Nikolaev Institute of Inorganic Chemistry, 630090 Novosibirsk, Russia}

\begin{abstract}

Lithium magnesium molybdate (Li$_2$Mg$_2$(MoO$_4$)$_3$) crystals
were grown by the low-thermal-gradient Czochralski  method.
Luminescence properties of the material (emission spectra,
thermally stimulated luminescence, dependence of intensity on
temperature, phosphorescence) have been studied under X-Ray
excitation in the temperature interval from 8 K to 400 K, while at
the same being operated as a scintillating bolometer at 20 mK for
the first time. We demonstrated that Li$_2$Mg$_2$(MoO$_4)_3$
crystals are a potentially promising detector material to search
for neutrinoless double beta decay of $^{100}$Mo.
\end{abstract}

 \end{frontmatter}

\section{Introduction}

Neutrinoless double beta ($0\nu2\beta$) decay is of essential
interest to particle physics since this process violates lepton
number and requires the neutrino to be a massive Majorana particle
\cite{Barrea:2012,Rodejohann:2012,Delloro:2016,Vergados:2016}.
Generally speaking, $0\nu2\beta$ decay can be mediated by many
other beyond the Standard Model effects
\cite{Deppisch:2012,Bilenky:2015}.

Significant efforts have been made to observe the $0\nu2\beta$
decay, however; the decay have yet to be observed (see the reviews
\cite{Elliott:2012,Giuliani:2012,Cremonesi:2014,Gomes:2015,Sarazin:2015}
and recent results \cite{EXO-200,NEMO-3,CUORE,GERDA,Gando:2016}).
The experiments have set limits on the half-life of this process
in the range of $T_{1/2} \sim 10^{24} - 10^{26}$ yr for various
nuclei, allowing us to restrict the effective Majorana neutrino
mass at the level of $\langle m_{\nu} \rangle \sim 0.1-0.8$~eV.
The broad range of the neutrino mass limits results from the
uncertainty of the nuclear matrix elements calculations
\cite{Engel:2017}.

The experimental sensitivity should be further improved to explore
the inverted hierarchy of the neutrino mass ($\langle
m_{\nu}\rangle\approx 0.02-0.05$ eV) corresponding to $T_{1/2}
\sim 10^{26} - 10^{27}$ yr even for the most promising nuclei
candidates \cite{Vergados:2016,Engel:2017}. Experiments able to
detect so rare $0\nu2\beta$ decays should operate hundreds of kg
of an isotope of interest, and have high detection efficiency and
energy resolution, and a very low (ideally zero) background. Low
temperature scintillating bolometers have a great potential to
realize large-scale high-sensitivity $0\nu2\beta$ decay
experiments with several nuclei. The isotope $^{100}$Mo is one of
the most promising candidates thanks to the high endpoint energy
of decay $Q_{2\beta}=3034.36(17)$ keV \cite{Wang:2017}, a high
isotopic abundance $\delta=9.744(65)\%$ \cite{Meija:2016} (and
possibility of enrichment in large amount by gas-centrifugation),
and a high decay probability
\cite{Rodryguez:2010,Simkovic:2013,Hyvarinen:2015,Barea:2015}.

There are several molybdate crystals have already been tested as
scintillating bolometers: ZnMoO$_4$
\cite{Beeman:2012a,Beeman:2012b,Armengaud:2017}, Li$_2$MoO$_4$
\cite{Armengaud:2017,Bekker:2016}, Li$_2$Zn$_2$(MoO$_4$)$_3$
\cite{Bashmakova:2009}, CaMoO$_4$ \cite{Kim:2015} with their main
properties reported in Table \ref{tab:prop}.

\clearpage
\begin{landscape}
\begin{center}
\begin{longtable}{|l|l|l|l|l|l|}
\caption{Properties of Li$_2$MoO$_4$, Li$_2$Mg$_2$(MoO$_4$)$_3$,
Li$_2$Zn$_2$(MoO$_4$)$_3$, CaMoO$_4$, and ZnMoO$_4$ crystal
scintillators. Wavelength of scintillation emission maximum is
denoted as $\lambda_{max}$,
$N_{vol}$ denotes number of Mo atoms per volume of the crystals.}\\

\hline
 Properties                 & Li$_2$MoO$_4$ \cite{Bekker:2016}  & Li$_2$Mg$_2$(MoO$_4$)$_3$ & Li$_2$Zn$_2$(MoO$_4$)$_3$ \cite{Bashmakova:2009}  & CaMoO$_4$ \cite{Annenkov:2008} & ZnMoO$_4$ \cite{Chernyak:2013} \\
 \hline
 Density (g/cm$^3$)         & $3.02-3.07$                       & 3.89 \cite{Penkova:1977}  & 4.38                                              & $4.2-4.3$             & 4.3               \\
  ~                         & ~                                 & 3.798 \cite{Sebastian:2003}&                                                  &                       &                  \\
  ~                         & ~                                 & 3.82 (present work)       &                                                   &                       &                  \\
 Melting point ($^\circ$C)  & 701                               & 1033 \cite{Penkova:1977}  & 890                                               & $1445-1480$           & 1003              \\
 Structural type            & Phenacite                         & Pnma \cite{Sebastian:2003,Klevtsova:1970,Solodovnikov:2009}& Orthorhombic     & Scheelite             & Triclinic, $P1$   \\
 Hardness (Mohs)            &                                   &                           &                                                   & $3.5-4$               & 3.5               \\
 $\lambda_{max}$ (nm)       & 590                               & 585 (present work)        & 610                                               & 520                   & 625               \\
 Refractive index           & 1.44                              &                           & 2.0                                               & 1.98                  & $1.89-1.96$       \\
 Hygroscopicity             & Weak                              & No                        & No                                                & No                    & No       \\
 $N_{vol}$ (atoms/cm$^3$)   & $1.06\times10^{22}$               & $1.27\times10^{22}$       & $1.27\times10^{22}$                               & $1.28\times10^{22}$   & $1.15\times10^{22}$ \\
 \hline

\end{longtable}
\end{center}
\label{tab:prop}
\end{landscape}

In addition to the technical requirements described above, a
crystal scintillator for a large-scale high-sensitivity
$0\nu2\beta$ decay experiment should meet some practical
requirements: low material cost, the capability for mass
production, possibility of enriched molybdenum recycling to
minimize loss of the enriched isotope, and to produce more
scintillation elements. In a case of large scale cryogenic
experiments (e.g., in the CUPID project \cite{CUPID,CUPID-RD}),
one should place a maximum number of the nuclei of interest in a
restricted cryostat volume. From this point of view, the
Li$_2$Mg$_2$(MoO$_4$)$_3$ crystal looks to be one of the most
appropriate materials for cryogenic experiments with $^{100}$Mo.
Absence of hygroscopicity (in contrast to weak hygroscopicity of
Li$_2$MoO$_4$) is also an advantage of the material.

In the present work, a Li$_2$Mg$_2$(MoO$_4$)$_3$ crystal has been
investigated as possible low temperature scintillator for double
beta decay experiments with $^{100}$Mo. Crystal growth of the
compound is reported in Sec. \ref{sec:cryst-growth}, the
luminescent properties of the material are summarized in Sec.
\ref{sec:lum}, and the first test of a Li$_2$Mg$_2$(MoO$_4$)$_3$
crystal sample as scintillating bolometer is described in Sec.
\ref{sec:cryotest}.

\section{Growth of Li$_2$Mg$_2$(MoO$_4$)$_3$ crystals}
\label{sec:cryst-growth}

\begin{figure}[htbp]
\begin{center}
\resizebox{0.509\textwidth}{!}{\includegraphics{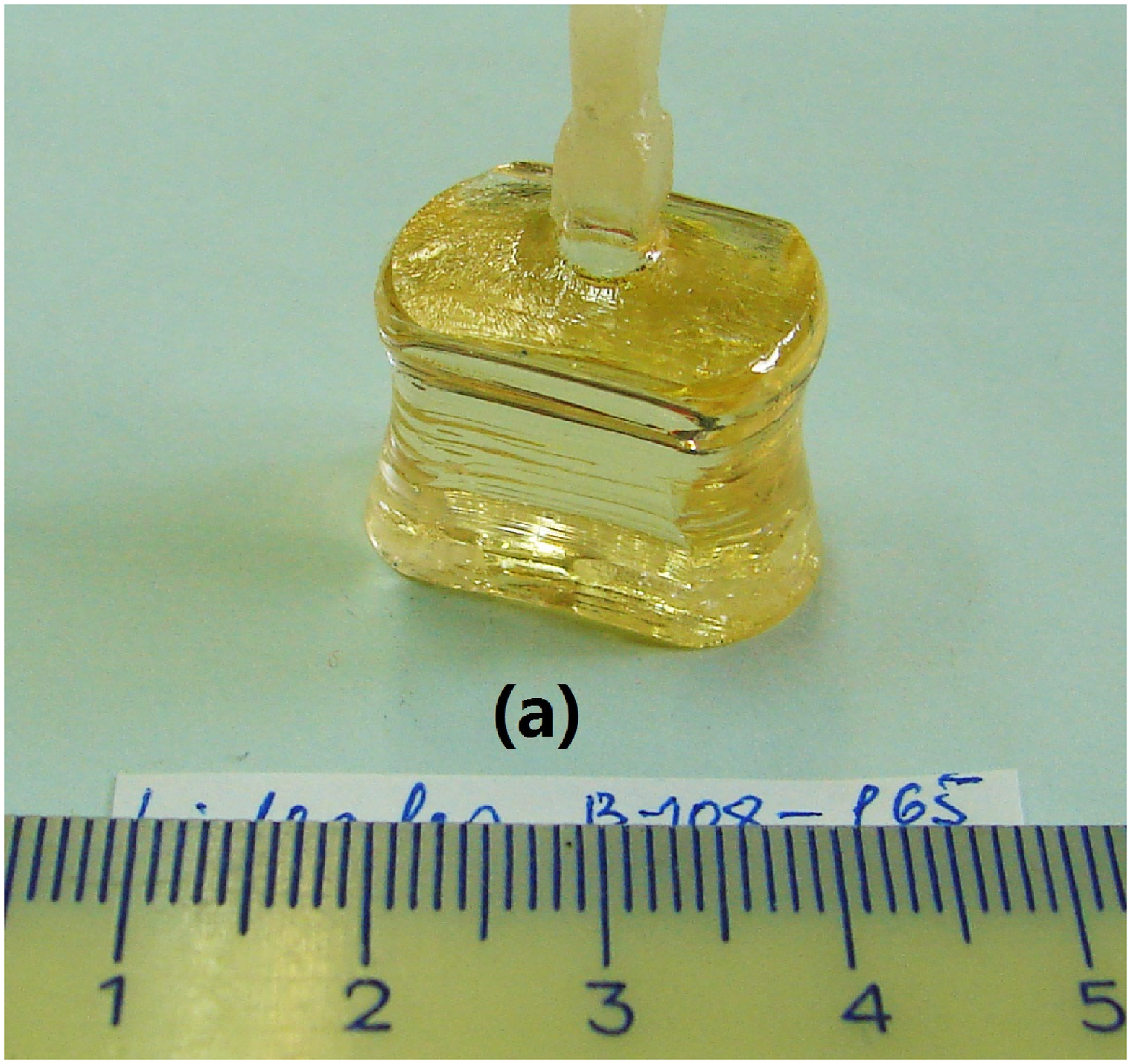}}
\resizebox{0.40\textwidth}{!}{\includegraphics{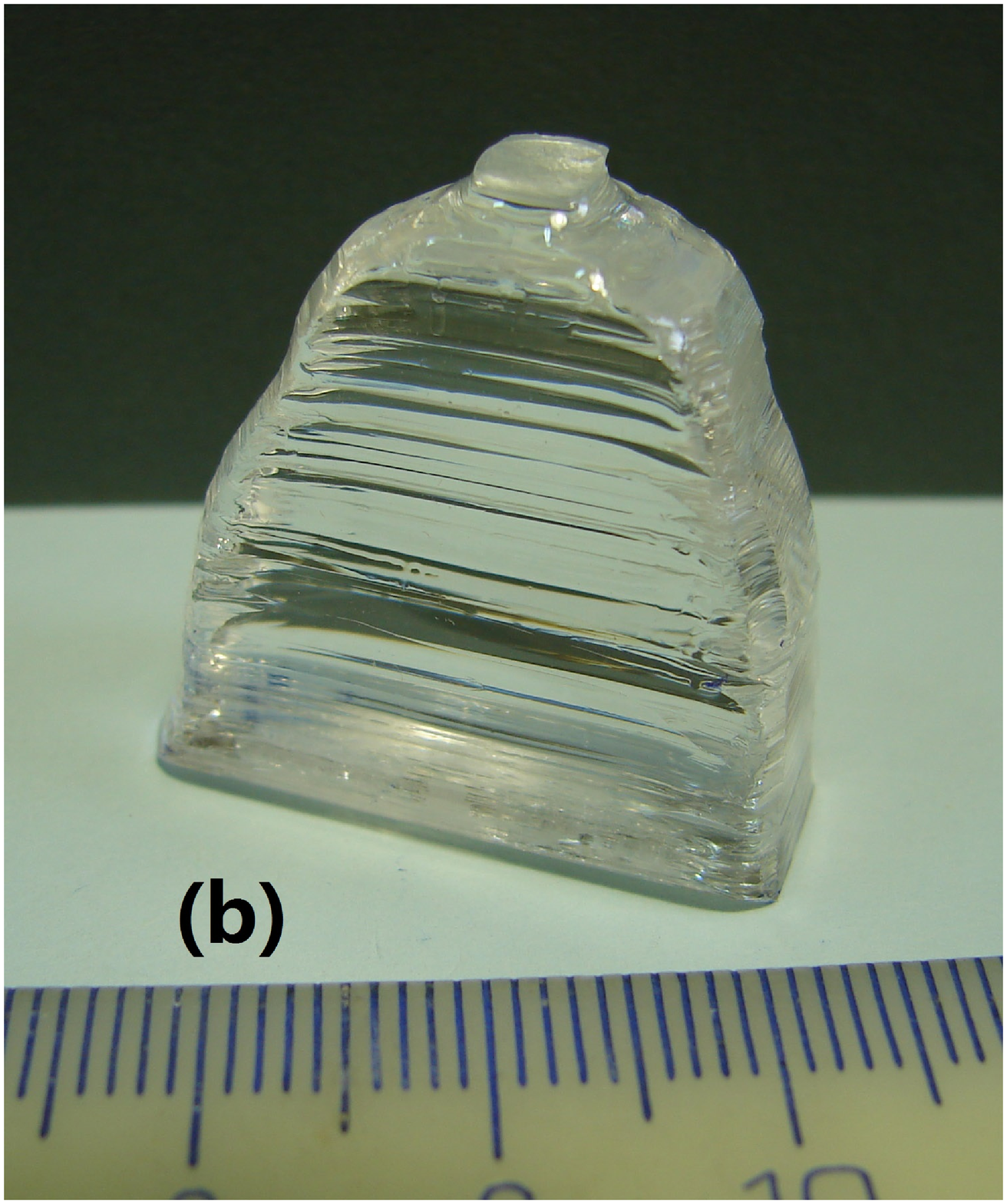}}
\caption{Boules of Li$_2$Mg$_2$(MoO$_4$)$_3$ crystals grown by the
low-thermal-gradient Czochralski method from 99.9\% (a) and
99.999\% (b) purity grade MoO$_3$. The improvement of the
crystal's optical quality thanks to the utilization of the higher
purity MoO$_3$ is clearly visible. The scales are in centimeters.}
 \label{fig:cryst}
\end{center}
\end{figure}

The main difficulty in the production of Li$_2$Mg$_2$(MoO$_4$)$_3$
crystals is the incongruent melting of the compound
\cite{Penkova:1977,Sebastian:2003}. This property makes it
difficult to grow large volume crystals by the ordinary
Czochralski  technique from a stoichiometric compound. In the
present study, Li$_2$Mg$_2$(MoO$_4$)$_3$ crystals were grown by
using the low-thermal-gradient Czochralski  method
\cite{Pavlyuk:1992,Trifonov:2013}. The Li$_2$Mg$_2$(MoO$_4$)$_3$
compound for the crystal growth has been synthesized by mixing
Li$_2$CO$_3$, MgO and MoO$_3$ powders in the ratio
Li$_2$Mg$_2$(MoO$_4)_3$ : Li$_2$MoO$_4=2:3$. In the first stage of
the crystal R\&D, a molybdenum oxide of 99.9\% purity grade has
been used, replaced by a 99.999\% high purity MoO$_3$ in later
development. The powders mixture was placed in a platinum crucible
$\oslash70 \times 120$ mm covered by a tight platinum lid with a
thin pipe, heated to a temperature $5-10$ degrees higher than the
melting temperature. The melt homogeneity was achieved mixing the
melt by a platinum roller unit over $2-3$ hours. The HX620 crystal
growing set-up utilized a three-zone resistance furnace with
weight monitoring. Several crystal boules have been grown from a
crystal seed of [010] orientation with a pulling rate $1-5$ mm per
day \cite{Patent:2011}. Two Li$_2$Mg$_2$(MoO$_4$)$_3$ crystal
boules produced from 99.9\% and 99.999\% purity MoO$_3$ are shown
in Fig. \ref{fig:cryst}.

 \nopagebreak

\begin{figure}[htb]
 \begin{center}
 \mbox{\epsfig{figure=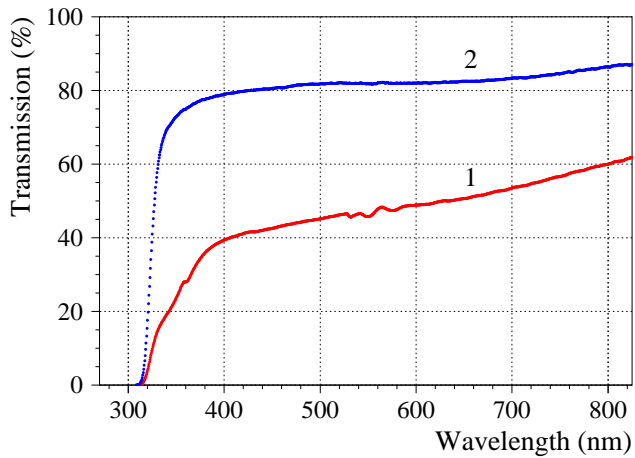,height=6.0cm}}
 \caption{(Color online) Optical transmission curves of 2 mm of Li$_2$Mg$_2$(MoO$_4$)$_3$
crystals produced from 99.9\% (1) and 99.999\% (2) purity grade
MoO$_3$.}
 \label{fig:opt}
 \end{center}
 \end{figure}

The higher optical quality of the Li$_2$Mg$_2$(MoO$_4$)$_3$
crystals produced from higher purity MoO$_3$ was confirmed by
transmittance measurements performed with a Shimadzu UV-3101PC
spectrometer. The results are presented in Fig.~\ref{fig:opt}. The
main cause of the crystals coloration is the contamination of the
raw materials (particularly of the molybdenum oxide) by transition
metals. The most abundant impurities of the 99.9\% purity
molybdenum oxide are K on the level of ~0.01 wt\%, and Si, Fe, Ti,
Ni, Ca, Cr (~0.001 wt\%). Presence of Fe, Ti, Ni, Cr can lead to
coloration of the crystal produced from the molybdenum oxide of
the 99.9\% purity grade. The effect of crystals coloration by
transition metals is well known, and was also observed in
molybdate crystals (see e.g. investigation of the raw-materials
purity level on the optical quality of zinc molybdate crystal
scintillators in \cite{Berge:2014}).

 \nopagebreak

Li$_2$Mg$_2$(MoO$_4$)$_3$ crystals are isotypic with lithium zinc
molybdate (Li$_2$Zn$_2$\\(MoO$_4$)$_3$) \cite{Solodovnikov:2009}.
Therefore, one can assume that Li$^+$ and Mg$^{2+}$ ions are
distributed over the $M1$, $M2$ and $M3$ sites (see Fig. 4 in
\cite{Solodovnikov:2009}). Taking into account that the
composition of Li$_2$Mg$_2$(MoO$_4$)$_3$ crystals may depend on
the synthesis temperature (as observed for
Li$_2$Zn$_2$(MoO$_4$)$_3$ \cite{Sebastian:2003}), one can expect
that the Li$_2$Mg$_2$(MoO$_4$)$_3$ crystal stoichiometry varies
due to the substitution/subtraction process 2Li$^{+}
\rightarrow$~Mg$^{2+}+~$vacancy. Therefore, the real crystal
composition is more accurately described by
Li$_{2-2x}$Mg$_{2+x}$(MoO$_4)_3$, where $0\leq  x \leq 0.3$.
However; the use of Li$_2$Mg$_2$(MoO$_4$)$_3$ compound in double
beta decay or dark matter experiments would require a large number
of identical crystals, with well-known compositions. It should be
stressed that the composition of the produced
Li$_2$Mg$_2$(MoO$_4$)$_3$ crystals is most strongly defined by the
composition of the powder used for the crystal growth. And
furthermore, it has been demonstrated in potassium gadolinium
tungstate (KGd(WO$_4$)$_2$, a similar compound from the point of
view of crystal growth), the low-thermal-gradient Czochralski
method allowed to produce large volume (up to 200 cm$^3$),
optically uniform KGd(WO$_4$)$_2$ crystals for laser applications
\cite{Mazur:2012}. R\&D is already underway to produce high
quality Li$_2$Mg$_2$(MoO$_4$)$_3$ crystal scintillators.

\section{Luminescence of Li$_2$Mg$_2$(MoO$_4$)$_3$ crystal under X-Ray irradiation}
\label{sec:lum}

\begin{figure}[htb]
 \begin{center}
 \mbox{\epsfig{figure=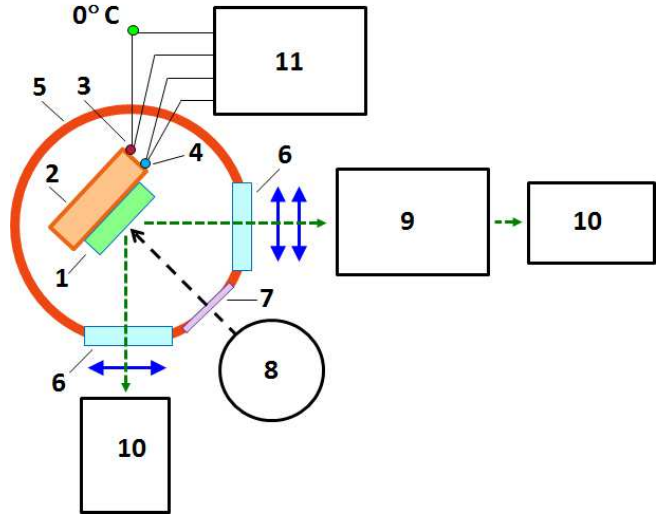,height=7.0cm}}
\caption{(Color online) Scheme of the set-up to study luminescence
of the Li$_2$Mg$_2$(MoO$_4$)$_3$ crystal under X-Ray irradiation:
(1) Li$_2$Mg$_2$(MoO$_4$)$_3$ crystal sample, (2) copper holder of
the crystal sample, (3) thermocouple, (4) semiconductor silicon
sensor, (5) vacuum cryostat, (6) quartz lens, (7) beryllium
window, (8) X-Ray tube, (9) monochromator, (10) photomultiplier
tube, (11) temperature measuring unit.}
 \label{fig:xrl-set-up}
 \end{center}
 \end{figure}

The luminescence of a Li$_2$Mg$_2$(MoO$_4$)$_3$ crystal sample
$10\times 2\times 2$ mm$^3$ was investigated under X-Ray
Irradiation (X-Ray Luminescence, XRL) in a wide temperature range
from 8 K to 400 K. The sample was mounted on a copper holder
inside a vacuum cryostat (a simplified scheme of the set-up is
presented in Fig. \ref{fig:xrl-set-up}). The temperature of the
copper holder was controlled by a semiconductor silicon sensor
WAD305 (in the temperature interval 8 K -- 85 K), and with a
chromel-copel thermocouple (in the temperature interval 85 K --
450 K). We showed in separate measurements, that the temperature
difference between a crystal sample and the support does not
exceed 0.2 K (0.4 K upon linear heating of the sample). The
crystal sample was irradiated through a beryllium window in the
cryostat by X-Rays from an X-Ray tube BHV-7 with a rhenium anode
operated at 20 kV, 25 mA with a flux of 0.635 mW/cm$^2$. The
luminescence was measured in two spectral channels: in a wide
wavelength interval (integral mode) and at a selected wavelength
548 nm (spectral mode). In the integral mode, the luminescence
light was focused with the help of a quartz lens on the
photocathode of a photomultiplier tube FEU-106 with extended
sensitivity in the wavelength region of $350-820$ nm. A
high-transmission monochromator MDR-2 (with a 600 mm$^{-1}$
diffraction grating) was used in the spectral mode. The emission
spectra were then corrected for the spectral sensitivity of the
registration system. The temperature dependencies of the XRL
intensity was obtained by cooling the sample to avoid effect of
Thermally Stimulated Luminescence (TSL). Phosphorescence
measurements were carried out after irradiation at temperatures of
8 K and 85 K over $\approx 10$ minutes (the exposure dose was
$\approx0.8$ J/cm$^2$). The TSL of the sample was measured after
the phosphorescence measurements while heating of the samples with
a rate $0.30\pm0.02$ K/s.

 \nopagebreak

\begin{figure}[htb]
 \begin{center}
 \mbox{\epsfig{figure=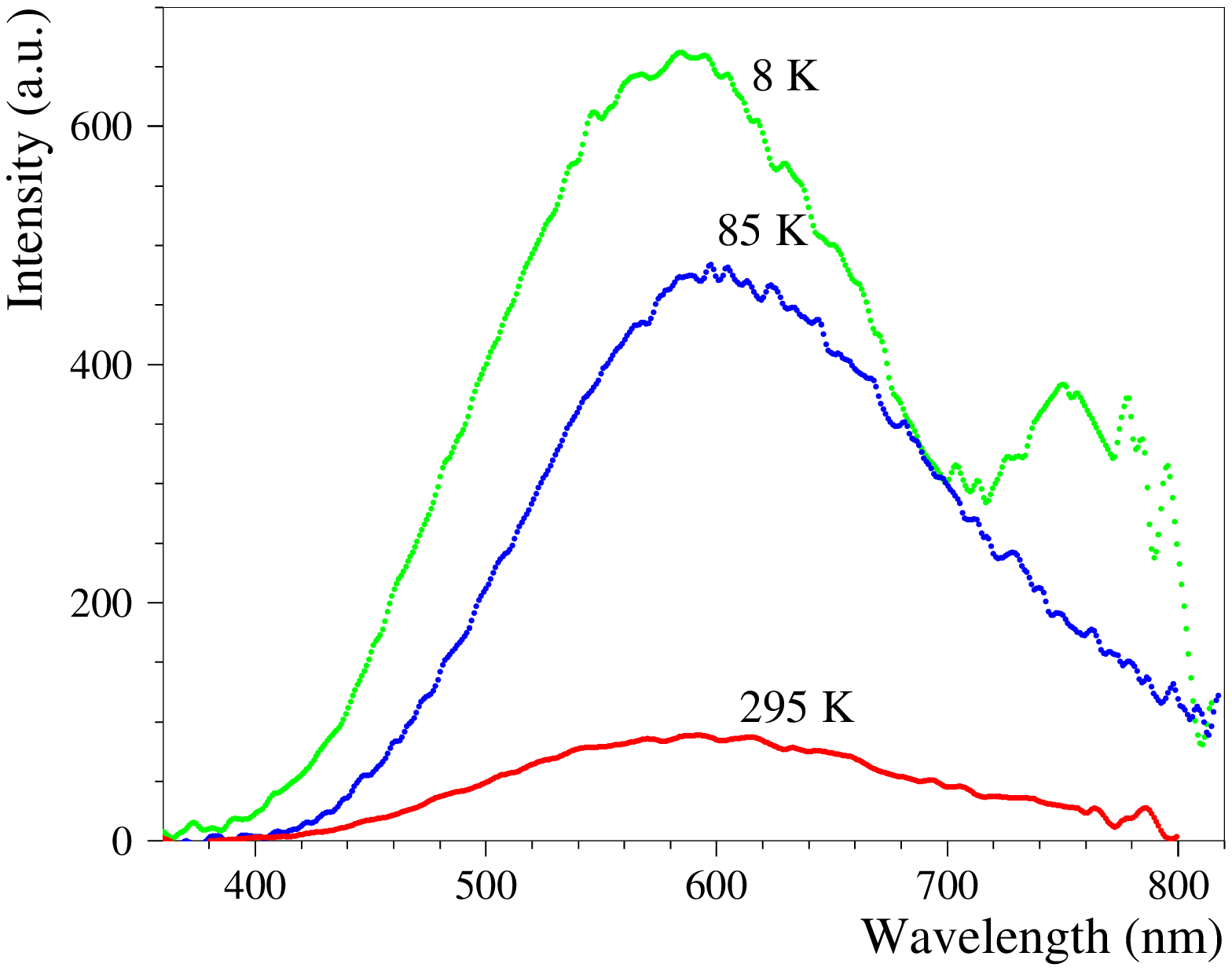,height=6.5cm}}
 \caption{(Color online) Emission spectra of Li$_2$Mg$_2$(MoO$_4$)$_3$ crystal under X-Ray irradiation at 8 K, 85 K and 295 K.}
 \label{fig:sp}
 \end{center}
 \end{figure}

The emission spectra of Li$_2$Mg$_2$(MoO$_4$)$_3$ crystal measured
under X-Ray irradiation at 8 K, 85 K and 295 K are shown in Fig.
\ref{fig:sp}. The main broad emission band with a maximum at
$\approx 600$ nm is observed at room and liquid nitrogen
temperatures, while the position of the maximum is slightly
shifted to a shorter wavelength $\approx 585$ nm at 8 K. An
additional infrared band appeared at $\sim750$ nm with cooling of
the sample to the liquid helium temperature. The measured emission
spectra are slightly different from those observed under
ultraviolet excitation with the maximum at 520 nm
\cite{Ryadun:2016}. It should be stressed that the luminescence
emission wavelength range of Li$_2$Mg$_2$(MoO$_4$)$_3$ crystal is
suitable for application of the material as low temperature
scintillating bolometer as discussed in Sec. \ref{sec:cryotest}.
Indeed, the wide spectral sensitivity of germanium photodetectors
($400-1700$ nm \cite{Larason:1998}) covers the emission spectral
range of the Li$_2$Mg$_2$(MoO$_4$)$_3$ crystal.

 \nopagebreak

\begin{figure}[htb]
 \begin{center}
 \mbox{\epsfig{figure=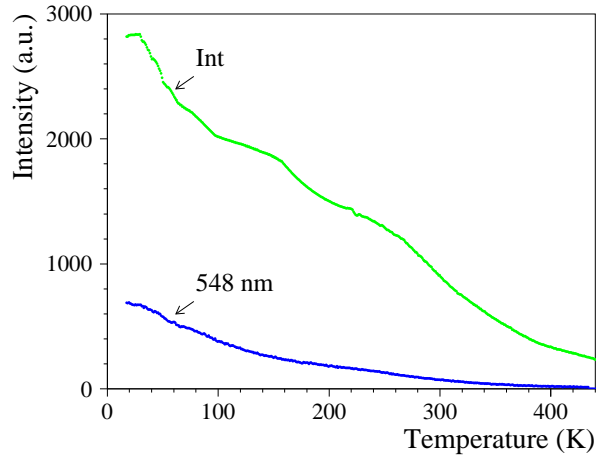,height=6.0cm}}
 \caption{(Color online) Temperature dependencies of XRL intensity of Li$_2$Mg$_2$(MoO$_4$)$_3$ crystal measured in the integral ("Int") and spectral (at 548 nm) modes of measurements.}
 \label{fig:int}
 \end{center}
 \end{figure}

The dependencies of the Li$_2$Mg$_2$(MoO$_4$)$_3$ luminescence
intensity on temperature measured in the integral and spectral
modes are presented in Fig. \ref{fig:int}. The luminescence
intensity increases by more than three times by cooling the
sample. Similar properties were observed in Li$_2$MoO$_4$
\cite{Bekker:2016} and Li$_2$Zn$_2$(MoO$_4$)$_3$
\cite{Bashmakova:2009}. Other molybdate crystals (as, e.g.,
CaMoO$_4$ \cite{Mikhailik:2007}, ZnMoO$_4$ \cite{Chernyak:2013},
SrMoO$_4$ \cite{Mikhailik:2015}) are also characterized by an
increasing of luminescence and scintillation efficiency as they
are cooled down to near liquid helium temperatures.

\clearpage
 \nopagebreak

\begin{figure}[htb]
 \begin{center}
 \mbox{\epsfig{figure=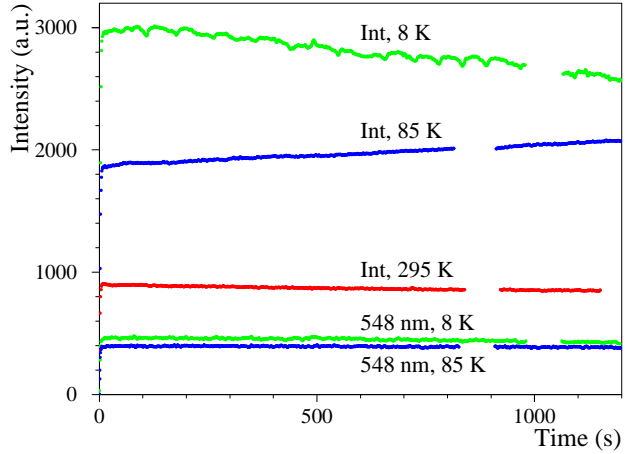,height=6.0cm}}
\caption{(Color online) Dependencies of Li$_2$Mg$_2$(MoO$_4$)$_3$
crystal XRL intensities on duration of X-Ray irradiation at 8 K,
85 K and 295 K. The data at 8 K and 85 K were recorded also in the
spectral mode at 548 nm.}
 \label{fig:dos}
 \end{center}
 \end{figure}

The luminescence intensity during X-Ray excitation in
Li$_2$Mg$_2$(MoO$_4$)$_3$ crystal is almost constant at room
temperature (see Fig. \ref{fig:dos}). A rather weak increase of
the XRL intensity at 77 K by $\approx 12$\% with a decrease at 8 K
by $\approx 13$\% after irradiation over 20 minutes were observed.
A systematic error of the measurements does not exceeds
$1.5\%-3\%$. It should be noted that after heating the sample up
to 450 K and then cooling again (without irradiation during the
cooling) to low temperatures (8 K or 85 K), the dose dependencies
of the XRL intensity remain the same. Taking into account the
constant XRL intensity at 295 K, one could assume that the change
in the XRL intensity at low temperatures is due to the processes
of charge carriers accumulation in deep traps and recharging of
recombination centers \cite{Degoda:2017}. This hypothesis is also
confirmed by the presence of phosphorescence and TSL described
below. In general, the observed independence of the
Li$_2$Mg$_2$(MoO$_4$)$_3$ luminescence on the dose (that is
proportional to the duration of X-Ray irradiation) indicates a
high radiation resistance of the material.

\clearpage
 \nopagebreak

\begin{figure}[htb]
 \begin{center}
 \mbox{\epsfig{figure=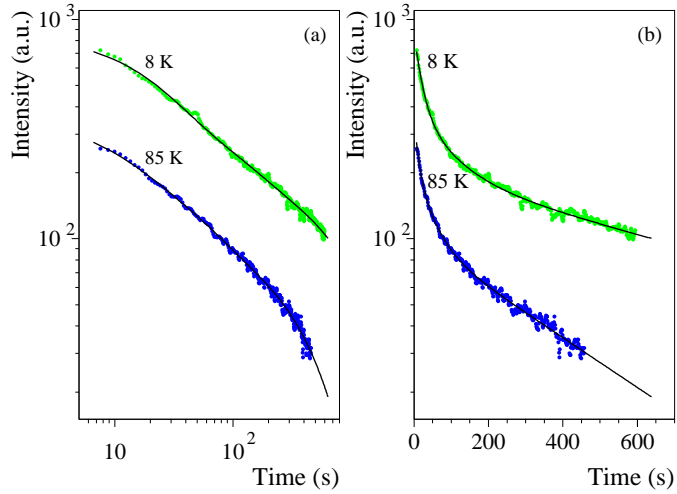,height=6.5cm}}
\caption{(Color online) The phosphorescence kinetic curves in
double logarithmic (a) and logarithmic (b) scales measured with
the Li$_2$Mg$_2$(MoO$_4$)$_3$ crystal after X-Ray irradiation at
temperatures 8 K and 85 K.}
 \label{fig:phosph}
 \end{center}
 \end{figure}

The Li$_2$Mg$_2$(MoO$_4$)$_3$ crystal showed a substantial
phosphorescence, as one can see in Fig. \ref{fig:phosph}, where
the phosphorescence decay curves recorded at 8 K and 85 K are
shown. As it was demonstrated for ZnMoO$_4$ crystals, long term
phosphorescence can be described by three exponential functions
better than by the Becquerel hyperbolic formula
\cite{Degoda:2017}. The fits of the Li$_2$Mg$_2$(MoO$_4$)$_3$
phosphorescence decay curves are presented in
Fig.~\ref{fig:phosph}. E.g., for the phosphorescence at 85 K the
decay times are $\tau_{1}=10(2)$ s, $\tau_{2}=42(5)$ s and
$\tau_{3}=380(10)$ s, while at 8 K the decay times are
$\tau_{1}=20(2)$ s, $\tau_{2}=91(12)$ s and $\tau_{3}=940(70)$ s.
The $\tau_{1}$ represents the effects of free charge recombination
coupled with charge carriers delocalized from shallow traps,
$\tau_2$ and $\tau_3$ stem from delocalization from phosphorescent
traps and deep traps, respectively. The phosphorescence shows the
recombination character of luminescence in
Li$_2$Mg$_2$(MoO$_4$)$_3$. Similar phosphorescence properties were
observed for ZnMoO$_4$ \cite{Degoda:2017} and Li$_2$MoO$_4$
\cite{Bekker:2016} crystals. The long term phosphorescence
indicates that the material tends to have some afterglow, a well
known effect in scintillators. However; the intensity of the
signal due to afterglow after $\sim0.1$ s of energy release in the
Li$_2$Mg$_2$(MoO$_4$)$_3$ crystal (the time to analyze
scintillation pulse profiles in low temperature scintillating
bolometers is typically shorter) decreases by two orders of
amplitude, approaching the photodetectors noise. Moreover; further
improvement of the crystals quality should reduce the
phosphorescence and possible afterglow effects.

 \nopagebreak

\begin{figure}[htb]
 \begin{center}
 \mbox{\epsfig{figure=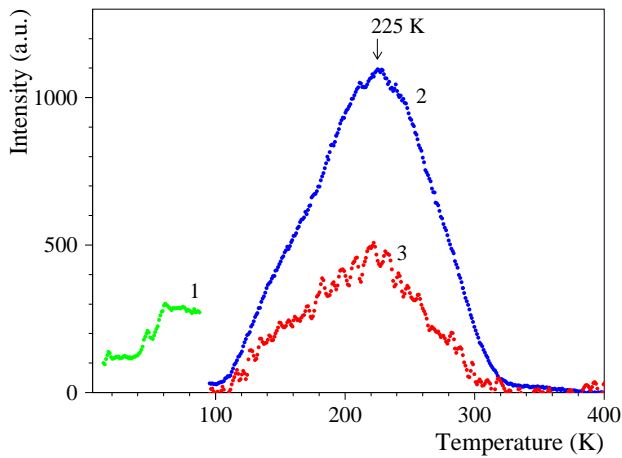,height=6.0cm}}
\caption{(Color online) Thermally stimulated luminescence of
Li$_2$Mg$_2$(MoO$_4$)$_3$ crystal measured after X-Ray irradiation
at temperature 8 K in integral mode (1), and at 85 K in integral
(2) and spectral (3) modes.}
 \label{fig:tsl}
 \end{center}
 \end{figure}

The TSL of Li$_2$Mg$_2$(MoO$_4$)$_3$ crystal measured after X-Ray
irradiation at 8 K and 85 K are presented in Fig. \ref{fig:tsl}.
It should be noted that use of two excitation temperatures allows
more accurate investigations of TSL, since accumulation of charge
carriers on the traps depends on the temperature of the
excitation. The most intensive TSL peak is observed at $\sim225$ K
(similar to Li$_2$Zn$_2$(MoO$_4$)$_3$ \cite{Bashmakova:2009}). At
room temperature, the traps responsible for the TSL become shallow
resulting in a stronger radiation resistance of the material (see
Fig. \ref{fig:dos}).

Both the intensive phosphorescence and the TSL prove the presence
of traps in the Li$_2$Mg$_2$(MoO$_4$)$_3$ crystal sample due to
defects, which indicates that there is still room to improve the
material.

\clearpage
 \nopagebreak

\section{Performance tests at milli-Kelvin temperature}
\label{sec:cryotest}

\begin{figure}[htbp]
\begin{center}
\resizebox{0.50\textwidth}{!}{\includegraphics{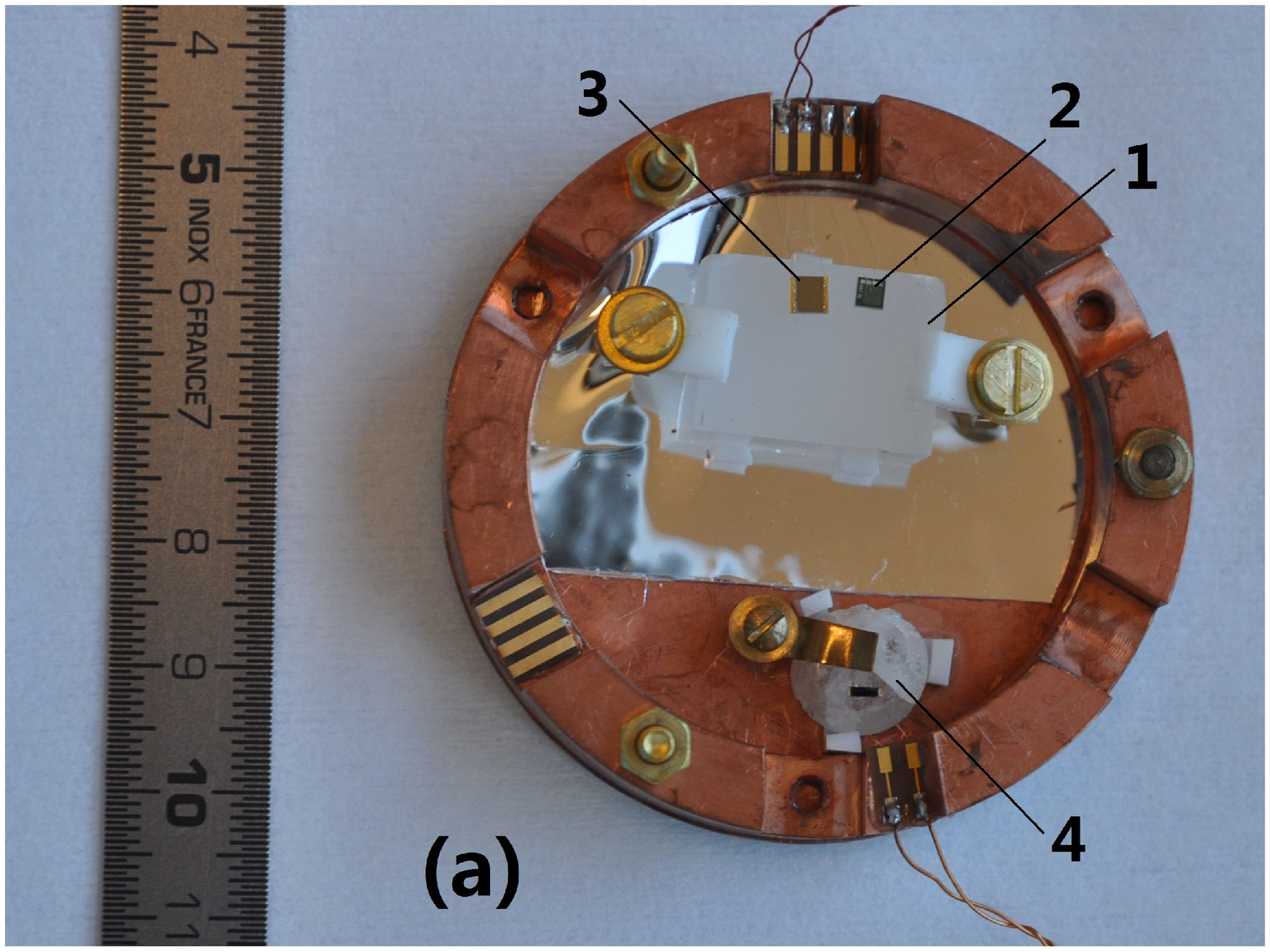}}
\resizebox{0.36\textwidth}{!}{\includegraphics{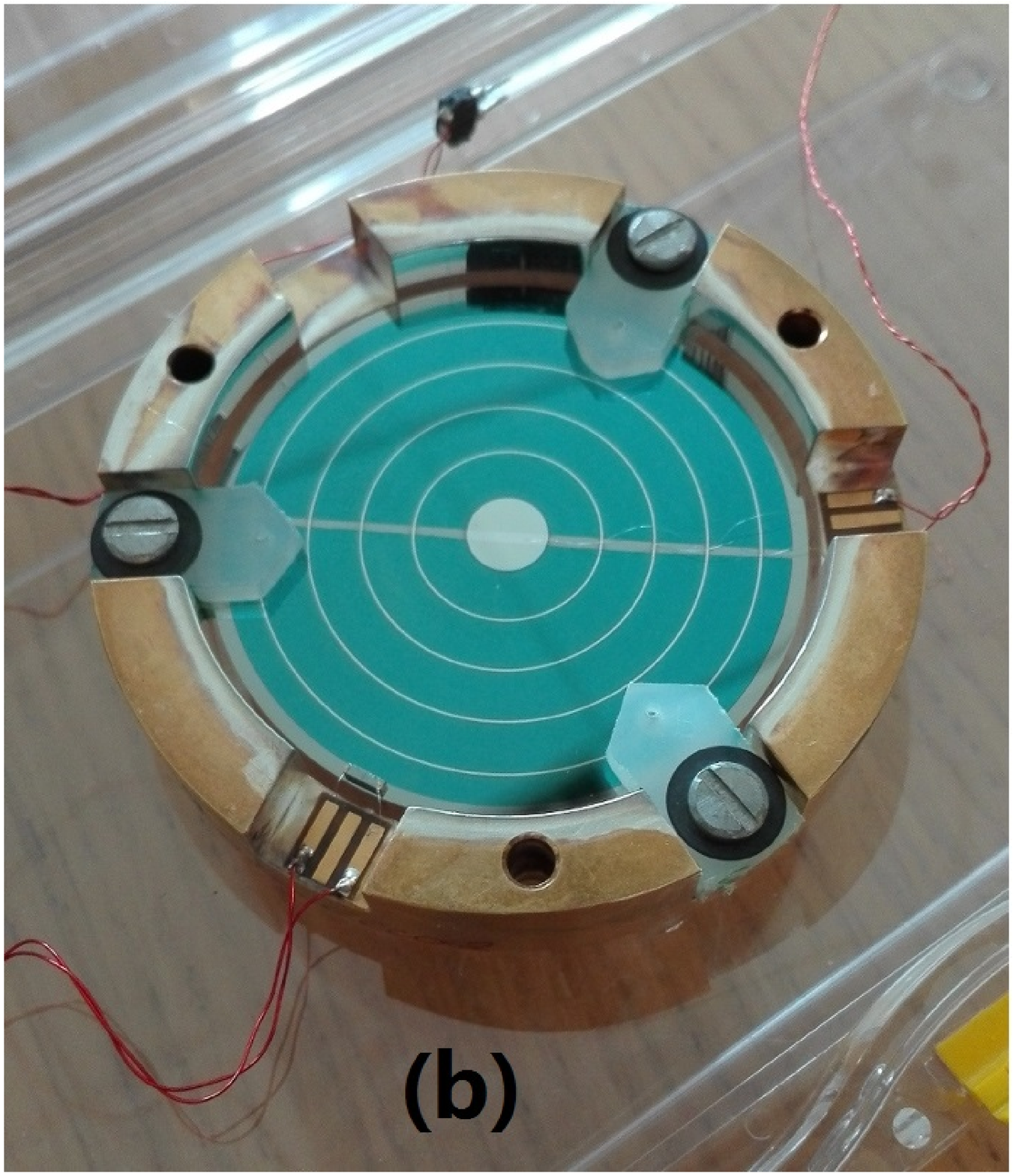}}
\caption{(Color online) (a) Li$_2$Mg$_2$(MoO$_4$)$_3$ crystal
sample (1) fixed on the copper frame and equipped with an NTD Ge
thermistor (2) and a silicon heater (3). (4) Another crystal
tested in the same set-up. (b) A Neganov-Luke Ge light detector
used to read out scintillation signals from the
Li$_2$Mg$_2$(MoO$_4$)$_3$ crystal.}
 \label{fig:setup}
\end{center}
\end{figure}

A Li$_2$Mg$_2$(MoO$_4$)$_3$ crystal sample (10.24 g, $1.9\times
1.4\times 1.0$ cm$^3$) was equipped with a Neutron Transmutation
Doped (NTD) Ge thermistor and a silicon heater (for
stabilization), which was than mounted in a copper frame (see Fig.
\ref{fig:setup} (a)) to run as bolometer.

Scintillation signals from the Li$_2$Mg$_2$(MoO$_4$)$_3$ crystal
were recorded with the help of a photodetector operated at 60 V in
Neganov-Luke mode \cite{Neganov:1981,Luke:1988}. The photodetector
(see Fig. \ref{fig:setup} (b)) consists of a germanium disk 44 mm
in diameter with a set of concentric annular Al electrodes with a
pitch 3.7 mm and coated with a SiO anti-reflective film 70 nm
thick (cyan color area on the picture). 60 V voltage was applied
between the electrodes to amplify the scintillation signals. Heat
signals after absorption of scintillation photons are read-out by
an NTD germanium thermistor (visible in the lower part of the
picture, attached at the germanium disk near the edge).

The detector module was run in a pulse-tube dilution refrigerator
\cite{Mancuso:2014}. The set-up is located aboveground at the
CSNSM (Orsay, France). A weak $^{210}$Po alpha source was
installed in the detector module to irradiate the crystal. The
cryostat is surrounded by a massive low-activity lead shield with
10 cm maximum thickness to reduce environmental gamma background
and the soft component of cosmic rays. The data were acquired in
stream mode with a 5 kHz sampling rate.

 \nopagebreak

\begin{figure}[htb]
 \begin{center}
 \mbox{\epsfig{figure=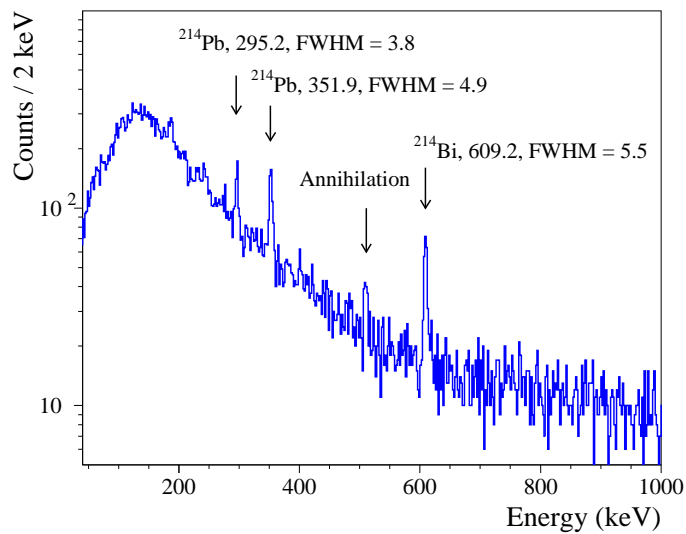,height=7.0cm}}
\caption{(Color online) Background energy spectrum accumulated
over 88 h with a 10 g Li$_2$Mg$_2$(MoO$_4$)$_3$ scintillating
bolometer at 20 mK. Energies of $\gamma$ quanta and values of the
energy resolution (FWHM) are in keV.}
 \label{fig:bg}
 \end{center}
 \end{figure}

A background energy spectrum was gathered over 88 h with the
Li$_2$Mg$_2$(MoO$_4$)$_3$ scintillating bolometer at 20 mK as
shown in in Fig. \ref{fig:bg}. Several gamma peaks (gamma quanta
of $^{214}$Pb and $^{214}$Bi, daughters of environmental
$^{226}$Ra) are visible in the spectrum which allowed us to
calibrate the detector and estimate its energy resolution (full
width at half maximum, FWHM) as shown in the Fig. \ref{fig:bg}.
For instance, the energy resolution of the detector was measured
as FWHM~$=5.5$ keV at 609 keV (gamma quanta of $^{214}$Bi), while
the baseline energy resolution of the detector module was 3.8 keV.
The spectrometric characteristics of a
Li$_2$Mg$_2$(MoO$_4$)$_3$-based detector could be further improved
operating the device in lower-level background conditions and at
lower temperature.

\clearpage
 \nopagebreak

\begin{figure}[htb]
 \begin{center}
 \mbox{\epsfig{figure=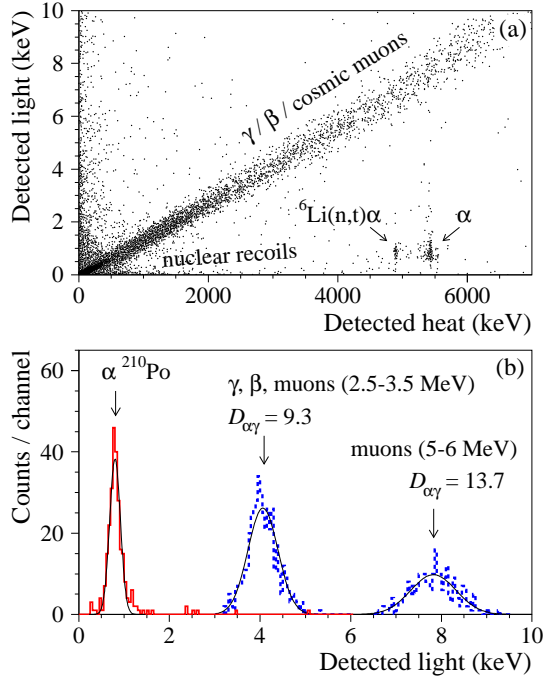,height=9.0cm}}
 \caption{(Color online) (a) Scatter plot of the light versus the heat signal
amplitudes accumulated with Li$_2$Mg$_2$(MoO$_4$)$_3$
scintillating bolometer over 88 h. The $\gamma$, $\beta$, muon
events are clearly separated from the neutron induced nuclear
recoils, $\alpha$-triton, and $\alpha$ events of $^{210}$Po. (b)
Discrimination between $\alpha$ events of $^{210}$Po (solid
histogram) and $\gamma$, $\beta$, cosmic muons events in the
energy intervals $2.5-3.5$ MeV and $5-6$ MeV (dashed histograms)
with the discrimination power $D_{\alpha\gamma}=9.3$ and
$D_{\alpha\gamma}=13.7$, respectively (see equation (\ref{eq:dp})
for $D_{\alpha\gamma}$ definition). The fits of the distributions
by Gaussian functions are shown by solid lines.}
 \label{fig:sc}
 \end{center}
 \end{figure}

The particle discrimination capability of the detector (an
important characteristic for application in rare-event searches)
is demonstrated by Fig. \ref{fig:sc} with data accumulated over 88
h with the Li$_2$Mg$_2$(MoO$_4)_3$ scintillating bolometer. The
$\gamma$, $\beta$, and cosmic muon events are clearly separated
from the $\alpha$-triton and $\alpha$ events\footnote{The events
between the bands can be explained by pulses pile-ups due to
comparatively slow response of bolometric detectors (the time
window of Li$_2$Mg$_2$(MoO$_4$)$_3$ heat pulses is $\sim0.07$ s).
The pile-up events nature was checked by their pulse profile
analysis.}.

\clearpage
 \nopagebreak

The particle discrimination capability (denoted here as
$D_{\alpha\gamma}$) to discriminate $\alpha$ particles and
$\gamma$ quanta ($\beta$ particles, muons) can be determined by
the following formula:

\begin{equation} \label{eq:dp}
 D_{\alpha\gamma} = |A_{\alpha}-A_{\gamma}|/
 \sqrt{\sigma^2_{\alpha}+\sigma^2_{\gamma}},
\end{equation}

\noindent where $A$ ($\sigma$) are average values (standard
deviations) of the scintillation signals distributions for
$\alpha$ particles and $\gamma$ quanta (or other particles such as
betas and muons). We obtained $D_{\alpha\gamma}=9.3(13.7)$ for 5.3
MeV $\alpha$ particles of $^{210}$Po and $\gamma$ ($\beta$, muons)
events in the energy intervals $2.5-3.5$ MeV ($5-6$ MeV). The
distributions are shown in Fig. \ref{fig:sc} (b). However, the
discrimination power should be estimated in the same energy
interval for $\alpha$ particles and $\gamma$ quanta ($\beta$
particles) in the energy region of the $0\nu2\beta$ peak of
$^{100}$Mo. Unfortunately, the statistic of $\alpha$ events with
energy $2.5-3.5$ MeV in our data is rather poor. We are going to
utilize an $\alpha$ source emitting energy-degraded $\alpha$
particles to investigate the discrimination power of
Li$_2$Mg$_2$(MoO$_4$)$_3$-based scintillating bolometers in our
further studies of the detector material. Nevertheless, the
achieved discrimination power is high enough to discriminate
$\beta$ and $\alpha$ events in a double beta experiment with
$^{100}$Mo.

\begin{figure}[htb]
 \begin{center}
 \mbox{\epsfig{figure=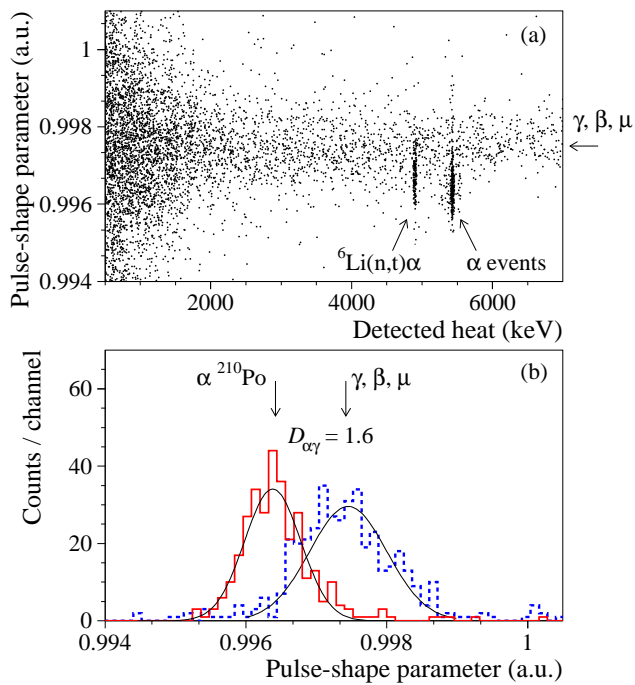,height=9.0cm}}
\caption{(Color online) (a) Scatter plot of the pulse-shape
parameter (see text) versus the heat signal amplitudes accumulated
with Li$_2$Mg$_2$(MoO$_4$)$_3$ scintillating bolometer over 88 h
of background run. (b) The positions of the pulse-shape parameter
distributions for $\alpha$ events of $^{210}$Po (solid histogram),
and for the $\gamma$, $\beta$, muon events with energy
$2.5-3.5$~MeV (dashed histogram) are slightly different due to the
difference in the heat signals shape. Fits of the distributions by
Gaussian functions are shown by solid lines.}
 \label{fig:scheat}
 \end{center}
 \end{figure}

Moreover; some differences were observed in the heat-pulse shapes
of $\gamma$ ($\beta$, cosmic muons) and $\alpha$ events. For this
analysis, we applied a parameter (pulse-shape parameter) that is a
ratio between the fitted amplitude (a value of signal maximum
obtained by the pulse fit) to the filtered amplitude (a signal
maximum after applying the optimum filter
\cite{Radeka:1967,Gatti:1986}). A background scatter plot of the
pulse-shape parameter versus the heat signal amplitudes
accumulated with the Li$_2$Mg$_2$(MoO$_4$)$_3$ scintillating
bolometer over 88 h is presented in Fig. \ref{fig:scheat} (a). The
distributions of the pulse-shape parameter for $\gamma$, $\beta$,
muon events with energy $2.5-3.5$~MeV, and $\alpha$ events of
$^{210}$Po are shown in Fig. \ref{fig:scheat} (b). The
discrimination power for the distributions calculated by equation
(\ref{eq:dp}) is $D_{\alpha\gamma}=1.6$.

 \nopagebreak

The scintillation light yield of the Li$_2$Mg$_2$(MoO$_4$)$_3$
crystal scintillator (the detected light energy per particle
energy measured by the deposited heat) is estimated to be 1.3
keV/MeV. The estimation was done thanks to the calibration of the
photodetector by using 5.9 keV X-rays from a weak $^{55}$Fe
source. The value is comparable to the light yield observed with
ZnMoO$_4$ ($0.8-1.5$ keV/MeV) and slightly exceeds the light yield
of Li$_2$MoO$_4$ crystal scintillators ($0.7-1.0$ keV/MeV)
\cite{Armengaud:2017}.

The quenching factor for $\alpha$ particles in the
Li$_2$Mg$_2$(MoO$_4$)$_3$ scintillator was estimated as the ratio
between the detected scintillation signals amplitudes for $\alpha$
particles of $^{210}$Po (in the energy interval around the
$\alpha$ peak $5216-5411$ keV) and the scintillation signals
amplitudes for cosmic muons ($\gamma$ quanta, $\beta$ particles)
in the same energy interval as 0.105(2).

\begin{figure}[htb]
 \begin{center}
 \mbox{\epsfig{figure=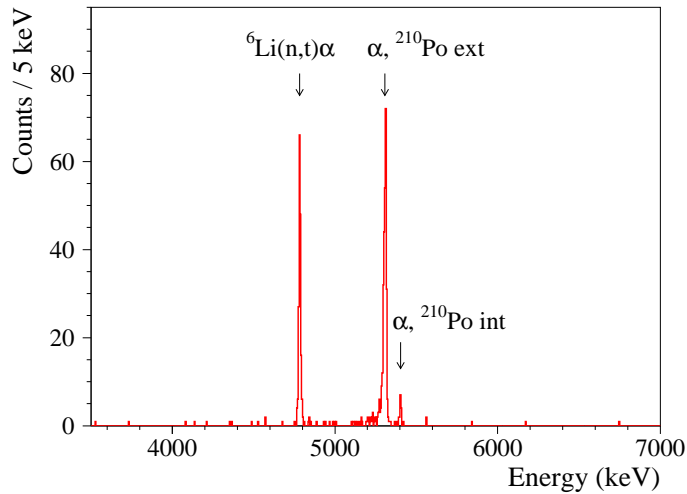,height=6.5cm}}
\caption{(Color online) Background energy spectrum of the
$\alpha$-triton and $\alpha$ events accumulated by the
Li$_2$Mg$_2$(MoO$_4$)$_3$ scintillating bolometer over 88 h.}
 \label{fig:alpha}
 \end{center}
 \end{figure}

An energy spectrum of the $\alpha$-triton and $\alpha$ events
accumulated by the Li$_2$Mg$_2$(MoO$_4$)$_3$ scintillating
bolometer was built by using the difference in scintillation yield
presented in Fig. \ref{fig:sc}. The obtained energy spectrum is
shown in Fig. \ref{fig:alpha}. There are two clearly visible
groups of events in the data: $\alpha$-triton peak with energy
4784 keV caused by thermal neutrons capture on $^6$Li, and two
peaks of $^{210}$Po. One peak is from the $^{210}$Po alpha source
that irradiated the crystal (the peak is labelled in Fig.
\ref{fig:alpha} as ``$^{210}$Po ext"), while the second weak peak
is due to the bulk contamination of the crystal by $^{210}$Po
(labelled on the Figure as ``$^{210}$Po int"). The energy
resolution is FWHM$~\approx12$ keV for the $\alpha$-triton peak,
and FWHM$~\approx9$ keV for the internal $^{210}$Po peak.

 \nopagebreak

Analysis of the alpha spectrum allowed us to evaluate radioactive
contamination of the Li$_2$Mg$_2$(MoO$_4$)$_3$ crystal. Activity
of the bulk $^{210}$Po can be estimated as $5.6(1.3)$ mBq/kg,
which is rather low taking into account that no particular efforts
were made to obtain radiopure material. It should be stressed that
we cannot conclude whether the $^{210}$Po is result of the crystal
contamination by $^{210}$Pb or it is $^{210}$Po itself. If it is
polonium contamination then the activity should decrease
substantially within a few years, due to the comparatively short
half-life of $^{210}$Po: $T_{1/2}\approx 138$ d. In contrast, the
activity of $^{210}$Po could even increase (see, e.g., Fig. 7 in
\cite{Danevich:2011}) in the case of led ($^{210}$Pb)
contamination (the half-life of $^{210}$Pb is $T_{1/2}\approx 22$
yr, the crystal was produced approximately 9 months before the
test). Due to the absence of other $\alpha$ peaks we were only
able to set limits ($\lim A$) on the active alpha members of the
$^{235}$U, $^{238}$U and $^{232}$Th chains with the equation
(\ref{eq:lim}):

\begin{equation}\label{eq:lim}
\lim A = \lim S/(\varepsilon\cdot\theta\cdot m\cdot t)
\end{equation}

\noindent where  $\varepsilon$ is the detection efficiency
(assumed to be 100\% due to a very short path length of alpha
particles in the Li$_2$Mg$_2$(MoO$_4$)$_3$ crystal), $\theta$ is
the $\alpha$ particle emission probability, $m$ is the mass of the
crystal sample, $t$ is the measuring time, and $\lim S$ is the
number of $\alpha$ events that can be excluded at a given
confidence level (C.L., all the limits here are given with 90\%
C.L.). Values of $\lim S$ were estimated by using the
Feldman-Cousins procedure for no effect observed on an estimated
background \cite{Feldman:1998}. The measured numbers of events
were calculated in the energy intervals $\pm 30$ keV around the
$Q_{\alpha}$ of the expected $\alpha$ peaks (the interval is wider
than $\pm$3 sigma for all expected $\alpha$ peaks, therefore, the
detection efficiency remains almost 100\%), while the background
counting rate was estimated in the energy intervals $3.3-4$ MeV
and $4.35-4.65$ MeV, where no $\alpha$ particles of U/Th are
expected. For instance, in the energy interval $4052-4112$ keV
(where a peak of $^{232}$Th with energy 4081.6 keV is expected) 1
event was detected, while the estimated background is 1.32 counts.
Therefore, according to \cite{Feldman:1998} we should exclude
$\lim S=3.06$ counts, that lead to the limit on $^{232}$Th
activity in the crystal $\leq 0.95$ mBq/kg. Other limits were
obtained in a similar way. A summary of the
Li$_2$Mg$_2$(MoO$_4$)$_3$ crystal radioactive contamination is
presented in Table~\ref{tab:rad-cont}.

\begin{table} [!htb]
\caption{Radioactive contamination of the
Li$_2$Mg$_2$(MoO$_4$)$_3$ crystal.}

\begin{center}
\begin{tabular}{|l|l|l|}

 \hline
 Chain      & Sub-chain     & Activity (mBq/kg) \\
 \hline
 $^{232}$Th & $^{232}$Th    & $\leq 0.95$       \\
 ~          & $^{228}$Th    & $\leq 1.1$        \\
 $^{238}$U  & $^{238}$U     & $\leq 0.95$       \\
 ~          & $^{234}$U     & $\leq 1.7$       \\
 ~          & $^{226}$Ra    & $\leq 1.7$        \\
 ~          & $^{210}$Po    & $5.6(1.3)$         \\
  $^{235}$U & $^{235}$U     & $\leq 1.9$        \\
  ~         & $^{223}$Ra    & $\leq 1.1$        \\
 \hline
\end{tabular}
\end{center}
\label{tab:rad-cont}
\end{table}

\section{Conclusions}

Large volume optically-clear quality Li$_2$Mg$_2$(MoO$_4$)$_3$
crystals were grown with the help of the low-thermal-gradient
Czochralski method. The luminescence of a
Li$_2$Mg$_2$(MoO$_4$)$_3$ crystal sample was studied under X-Ray
excitation in the $8-400$ K temperature range. Luminescence with a
maximum at $585$~nm is observed at 8~K. The luminescence intensity
increases by more than a factor 3 when cooling the crystal sample
from room temperature to 8 K. Intense phosphorescence and
thermally stimulated luminescence indicate presence of traps due
to defects in the crystal sample, which points to that fact that
the material quality can be further improved.

Low temperature measurements (at 20 mK) of a 10 g sample of
Li$_2$Mg$_2$(MoO$_4$)$_3$ scintillating bolometer were carried out
over 88 h, demonstrating energy resolution as good as 5.5 keV
(FWHM, at 609 keV), despite the comparatively high baseline noise
(3.8 keV, whereas it is typically on the level of $1-2$ keV). The
scintillator light yield is estimated as $\approx1.3$ keV/MeV. The
Li$_2$Mg$_2$(MoO$_4$)$_3$ scintillating bolometer showed an
excellent particle discrimination capability with a discrimination
power $D_{\alpha\gamma}=9$ (see equation \ref{eq:dp}) for
$2.5-3.5$ MeV $\gamma$ ($\beta$, muons) events and 5.3 MeV
$\alpha$ particles of $^{210}$Po. We have observed also a weak
difference in the heat signal shapes of $\gamma$ ($\beta$, cosmic
muons) and $\alpha$ particles. Despite no special efforts to
obtain radiopure material, the radioactive contamination of the
crystal was measured to be quite low. The activity of $^{210}$Po
is $\sim6$ mBq/kg, while only limits on the $\sim$ mBq/kg level
were estimated for other $\alpha$ active members of the $^{235}$U,
$^{238}$U and $^{232}$Th families. These bolometric measurements
have demonstrated that Li$_2$Mg$_2$(MoO$_4)_3$ is a potentially
promising detector material for double beta decay experiments with
molybdenum as well as other rare-event searches.

In particular, we would like to stress that
Li$_2$Mg$_2$(MoO$_4)_3$ scintillator contains a variety of
elements with different atomic masses $A\approx7$, 16, 24, 96 (Li,
O, Mg and Mo, respectively) which can be exploited in dark matter
low temperature scintillating bolometers (see CRESST
\cite{Angloher:2016}) to probe new areas of parameter space,
particularly of spin-dependent dark matter.

\section{Acknowledgements}

These studies were supported in part by the project
``Investigation of neutrino and weak interaction in double beta
decay of $^{100}$Mo" in the framework of the Programme ``Dnipro"
based on Ukraine-France Agreement on Cultural, Scientific and
Technological Cooperation, and by the IDEATE International
Associated Laboratory (LIA). F.A.~Danevich gratefully acknowledges
the support from the ``Jean d'Alembert" Grants program (Project
CYGNUS) of the University of Paris-Saclay. A.S.~Zolotarova is
supported by the ``IDI 2015'' project funded by the IDEX Paris-
Saclay, ANR-11-IDEX-0003-02. The authors are grateful to Alexander
Leder from the Massachusetts Institute of Technology for a careful
reading of the manuscript and helpful corrections.


\begin{thebibliography}{99}

 \bibitem{Barrea:2012} J.~Barea, J.~Kotila, F.~Iachello, Limits on Neutrino Masses from Neutrinoless Double-$\beta$ Decay, Phys. Rev. Lett. 109 (2012) 042501.
 \bibitem{Rodejohann:2012} W.~Rodejohann, Neutrino-less double beta decay and particle physics, J. Phys. G 39 (2012) 124008.
 \bibitem{Delloro:2016} S.~Dell'Oro, S.~Marcocci, M.~Viel, F.~Vissani, Neutrinoless Double Beta Decay: 2015 Review, AHEP 2016 (2016) 2162659.
 \bibitem{Vergados:2016} J.D.~Vergados, H.~Ejiri, F.~\v{S}imkovic, Neutrinoless double beta decay and neutrino mass, Int. J. Mod. Phys. E 25 (2016) 1630007.
 \bibitem{Deppisch:2012} F.F.~Deppisch, M.~Hirsch, H. P$\mathrm{\ddot{a}}$s, Neutrinoless double-beta decay and physics beyond the standard model, J. Phys. G 39 (2012) 124007.
 \bibitem{Bilenky:2015}  S.M.~Bilenky, C.~Giunti, Neutrinoless double-beta decay: A probe of physics beyond the Standard Model, Int. J. Mod. Phys. A 30 (2015) 1530001.
 \bibitem{Elliott:2012} S.R.~Elliott, Recent progress in double beta decay, Mod. Phys. Lett. A 27 (2012) 123009.
 \bibitem{Giuliani:2012} A.~Giuliani, A.~Poves, Neutrinoless Double-Beta Decay, AHEP 2012 (2012) 857016.
 \bibitem{Cremonesi:2014} O.~Cremonesi, M.~Pavan, Challenges in Double Beta Decay, AHEP 2014 (2014) 951432.
 \bibitem{Gomes:2015} J.J.~G$\mathrm{\acute{o}}$mez-Cadenas, J.~Mart$\mathrm{\acute{i}}$n-Albo, Phenomenology of Neutrinoless Double Beta Decay, Proc. of Sci. (GSSI14) 004 (2015) 1.
 \bibitem{Sarazin:2015} X.~Sarazin, Review of Double Beta Experiments, J. Phys.: Conf. Ser. 593 (2015) 012006.
 \bibitem{EXO-200} J.B.~Albert et al. (The EXO-200 Collaboration), Search for Majorana neutrinos with the first two years of EXO-200 data, Nature 510 (2014) 229.
 \bibitem{NEMO-3} R. Arnold et al., Results of the search for neutrinoless double-$\beta$ decay in $^{100}$Mo with the NEMO-3 experiment, Phys. Rev. D 92 (2015) 072011.
 \bibitem{CUORE} K. Alfonso et al. (CUORE Collaboration), Search for Neutrinoless Double-Beta Decay of $^{130}$Te with CUORE-0, Phys. Rev. Lett. 115 (2015) 102502.
 \bibitem{GERDA} M.~Agostini et al., Background-free search for neutrinoless double-$\beta$ decay of $^{76}$Ge with GERDA, Nature 544 (2017) 47.
 \bibitem{Gando:2016} A.~Gando et al. (KamLAND-Zen Collaboration), Search for Majorana Neutrinos Near the Inverted Mass Hierarchy Region with KamLAND-Zen, Phys. Rev. Lett. 117 (2016) 082503.
 \bibitem{Engel:2017} J.~Engel, J.~Men\'{e}ndez, Status and future of nuclear matrix elements for neutrinoless double-beta decay: a review, Rep. Prog. Phys. 80 (2017) 046301.
 \bibitem{Wang:2017} M.~Wang et al., The AME2016 atomic mass evaluation, Chin. Phys. C 41 (2017) 030003.
 \bibitem{Meija:2016} J.~Meija et al., Isotopic compositions of the elements 2013 (IUPAC Technical Report), Pure Appl. Chem. 88 (2016) 293.
 \bibitem{Rodryguez:2010} T.R.~Rodryguez, G.~Martynez-Pinedo, Energy Density Functional Study of Nuclear Matrix Elements for Neutrinoless $\beta\beta$ Decay, Phys. Rev. Lett. 105 (2010) 252503.
 \bibitem{Simkovic:2013} F.~~\v{S}imkovic, V.~Rodin, A.~Faessler, P.~Vogel, $0\nu\beta\beta$ and $2\nu\beta\beta$ nuclear matrix elements, quasiparticle random-phase approximation, and isospin symmetry restoration, Phys. Rev. C 87 (2013) 045501.
 \bibitem{Hyvarinen:2015} J.~Hyvarinen, J.~Suhonen, Nuclear matrix elements for $0\nu2\beta$  decays with light or heavy Majorana-neutrino exchange, Phys. Rev. C 91 (2015) 024613.
 \bibitem{Barea:2015} J.~Barea, J.~Kotila, F.~Iachello, $0\nu2\beta$ and $2\nu2\beta$ nuclear matrix elements in the interacting boson model with isospin restoration, Phys. Rev. C 91 (2015) 034304.
 \bibitem{Beeman:2012a} J.W.~Beeman et al., A next-generation neutrinoless double beta decay experiment based on ZnMoO$_4$ scintillating bolometers, Phys. Lett. B 710 (2012) 318.
 \bibitem{Beeman:2012b} J.W.~Beeman et al., ZnMoO$_4$: A promising bolometer for neutrinoless double beta decay searches, Astropart. Phys. 35 (2012) 813.
 \bibitem{Armengaud:2017} E.~Armengaud et al., Development of $^{100}$Mo-containing scintillating bolometers for a high-sensitivity neutrinoless double-beta decay search, Eur. Phys. J. C 77 (2017) 785.
 \bibitem{Bekker:2016} T.B.~Bekker et al., Aboveground test of an advanced Li$_2$MoO$_4$ scintillating bolometer to search for neutrinoless double beta decay of $^{100}$Mo, Astropart. Phys. 72 (2016) 38.
 \bibitem{Bashmakova:2009} N.V.~Bashmakova et al., Li$_2$Zn$_2$(MoO$_4)_3$ as a potential detector for $^{100}$Mo $2\beta$ search, Functional Materials 16 (2009) 266.
 \bibitem{Kim:2015} G.B.~Kim et al., A CaMoO$_4$ Crystal Low Temperature Detector for the AMoRE Neutrinoless Double Beta Decay Search, AHEP 2015 (2015) 817530.
 \bibitem{Annenkov:2008} A.N.~Annenkov et al., Development of CaMoO$_4$ crystal scintillators for a double beta decay experiment with $^{100}$Mo, Nucl. Instrum. Meth. A 584 (2008) 334.
 \bibitem{Chernyak:2013} D.M. Chernyak et al., Optical, luminescence and thermal properties of radiopure ZnMoO$_4$ crystals used in scintillating bolometers for double beta decay search, Nucl. Instrum. Meth. A 729 (2013) 856.
 \bibitem{Penkova:1977} V.G.~Penkova, P.V.~Klevtsov, Synthesis of crystals of double lithium molybdates with Mg, Ni, Co, Fe and Zn divalent metals, J. Inorg. Chem. 22 (1977) 1713 (in Russian).
 \bibitem{Sebastian:2003} L.~Sebastian et al., Synthesis, structure and lithium-ion conductivity of Li$_{2-2x}$Mg$_{2+x}$(MoO$_4)_3$ and Li$_3$M(MoO$_4)_3$ (M$^\mathrm{III}=$~Cr, Fe), J. Mater. Chem. 13 (2003) 1797.
 \bibitem{Klevtsova:1970} R.F.~Klevtsova, S.A.~Magarill, Crystal Structure of lithium iron molybdates Li$_3$Fe$^{\bullet\bullet\bullet}$(MoO$_4)_3$ and Li$_3$Fe$_2$$^{\bullet\bullet}$(MoO$_4)_3$, Crystallography Rep. 15 (1970) 710 (in Russian).
 \bibitem{Solodovnikov:2009} S.F.~Solodovnikov et al., Revised phase diagram of Li$_2$MoO$_4$--ZnMoO$_4$ system, crystal structure and crystal growth of lithium zinc molybdate, J. Solid State Chem. 182 (2009) 1935.
 \bibitem{CUPID} G. Wang et al., CUPID: CUORE (Cryogenic Underground Observatory for Rare Events) Upgrade with Particle IDentification, arXiv:1504.03599v1 [physics.ins-det].
 \bibitem{CUPID-RD} G. Wang et al., R\&D towards CUPID (CUORE Upgrade with Particle IDentification), arXiv:1504.03612v1 [physics.ins-det].
 \bibitem{Pavlyuk:1992} A.A.~Pavlyuk et al., Low Thermal Gradient technique and method for large oxide crystals growth from melt and flux, in the proceedings of the APSAM-92 (Asia Pacific Society for Advanced Materials), Shanghai, China, 26-29 April 1992, p. 164.
 \bibitem{Trifonov:2013} V.A.~Trifonov et al., Growth and spectroscopic characteristics of Li$_2$Mg$_2$(MoO$_4$)$_3$ and Li$_2$Mg$_2$(MoO$_4$)$_3$:Co$^{2+}$ crystals, Inorg. Mater. 49 (2013) 517.
 \bibitem{Patent:2011} A.A.~Pavlyuk, V.A.~Trifonov, An approach to grow lithium magnesium molybdate, Patent No. 2487968, 2011, Russian Federation.
 \bibitem{Berge:2014} L.~Berg\'{e} et al., Purification of molybdenum, growth and characterization of medium volume ZnMoO4 crystals for the LUMINEU program, JINST 9  (2014) P06004.
 \bibitem{Mazur:2012} M.M~Mazur et al., Elastic and photoelastic properties of KGd(WO$_4$)$_2$ single crystals, Akusticheskiy zhurnal 58 (2012) 701 (in Russian).
 \bibitem{Ryadun:2016} A.A.~Ryadun et al.,  Structre and properties of Li$_{2-2x}$Mg$_{2+x}$(MoO$_4)_3$ crystals activated by copper ions, J. Struct. Chem. 57 (2016) 459.
 \bibitem{Larason:1998} T.C.~Larason and S.S.~Bruce,  Spatial uniformity of responsivity for silicon, gallium nitride, germanium, and indium gallium arsenide photodiodes, Metrologia 35 (1998) 491.
 \bibitem{Mikhailik:2007} V.B.~Mikhailik, S.~Henry, H.~Kraus, I.~Solskii, Temperature dependence of CaMoO$_4$ scintillation properties, Nucl. Instrum. Meth. A 583 (2007) 350.
 \bibitem{Mikhailik:2015} V.B.~Mikhailik et al., Temperature dependence of scintillation properties of SrMoO$_4$, Nucl. Instrum. Meth. A 792 (2015) 1.
 \bibitem{Degoda:2017} V.Ya.~Degoda et al., Long time phosphorescence in ZnMoO$_4$ crystals, J. Luminescence 181 (2017) 269.
 \bibitem{Neganov:1981} B.S.~Neganov, V.N.~Trofimov, USSR Patent No. 1037771 (1981).
 \bibitem{Luke:1988} P.N.~Luke, Voltage-assisted calorimetric ionization detector, J. Appl. Phys. 64 (1988) 6858.
 \bibitem{Mancuso:2014} M.~Mancuso et al., An aboveground pulse-tube-based bolometric test facility for the validation of the LUMINEU ZnMoO$_4$ crystals, J. Low Temp. Phys. 176 (2014) 571.
 \bibitem{Radeka:1967} V.~Radeka, N.~Karlovac, Least-square-error amplitude measurement of pulse signals in presence of noise, Nucl. Instrum. Meth. 52 (1967) 86.
 \bibitem{Gatti:1986} E.~Gatti, P.F.~Manfredi, Processing the signals from solid-state detectors in elementary-particle physics, Riv. Nuovo Cimento 9 (1986) 1.
 \bibitem{Danevich:2011} F.A.~Danevich et al., Effect of recrystallisation on the radioactive contamination of CaWO$_4$ crystal scintillators, Nucl. Instrum. Meth. 631 (2011) 44.
 \bibitem{Feldman:1998} G.J.~Feldman, R.D.~Cousins, Unified approach to the classical statistical analysis of small signals, Phys. Rev. D 57 (1998) 3873.
 \bibitem{Angloher:2016} G.~Angloher et al., Results on light dark matter particles with a low-threshold CRESST-II detector, Eur. Phys. J. C 76 (2016) 25.

\end{thebibliography}
\end{document}